\documentclass[12pt,preprint]{aastex}

\newcommand{\um}{$\mu$m}

\shortauthors{Testi et al.}

\begin{document}

\title{A Young Very Low-Mass Object surrounded by warm dust}


\author{L. Testi\altaffilmark{1}, A. Natta\altaffilmark{1}, E. Oliva\altaffilmark{1,2}, F. D'Antona\altaffilmark{3}, F. Comer\'on\altaffilmark{4}, C. Baffa\altaffilmark{1}, G. Comoretto\altaffilmark{1} and S. Gennari\altaffilmark{1}}

\altaffiltext{1}{Osservatorio Astrofisico di Arcetri, INAF,
Largo E. Fermi 5, I-50125 Firenze, Italy}
\altaffiltext{2}{Telescopio Nazionale Galileo and Centro Galileo Galilei, INAF,
P.O. Box 565, E-38700, Santa Cruz de La Palma, Spain}
\altaffiltext{3}{Osservatorio Astronomico di Roma, INAF,
via Frascati 33, I-00044 Roma, Italy}
\altaffiltext{4}{European Southern Observatory, Karl-Schwarzschild-Strasse 2,
D-85748 Garching bei M\"unchen, Germany}

\begin{abstract}
We present a complete  low-resolution (R$\sim$100) near-infrared spectrum
of the substellar object GY~11, member of the $\rho$-Ophiuchi young association.
The object is remarkable because of its low estimated mass and age and  because it is
associated with a mid-infrared source, an indication of a surrounding 
dusty disk. Based on the comparison of our spectrum with similar spectra
of field M-dwarfs and  atmospheric models,
we obtain revised estimates of the
spectral type, effective temperature and luminosity of the central object.
These parameters are used to place the object on a Hertzprung-Russell diagram
and to compare with the prediction of pre-main sequence evolutionary models.
Our analysis suggests that the central object has a 
very low mass, probably below the deuterium burning limit and in the 
range 8--12~M$_{Jupiter}$, and a young age, less than 1~Myr.
The infrared excess is shown to be consistent with the emission of a flared,
irradiated disk
similar to those found in more massive brown dwarf and TTauri
systems. This result suggests that substellar objects,
even the so-called isolated planetary
mass objects, found in young stellar associations are produced in a similar
fashion as  stars, by core contraction and gravitational
collapse.
\end{abstract}



\section{Introduction}
\label{sintro}

The discovery 
of Brown Dwarfs (BDs) and
objects with  masses comparable to those of giant planets, well
below the deuterium burning limit
(M$<$13~M$_J$), 
``free-floating" in young stellar clusters \citep{Zea00,LR00}
has opened an interesting debate on their origin.
Do they  form like ordinary stars from the collapse of
molecular cores \citep{SAL87}? If so, the very existence of 
very low-mass objects
and their mass function put strong constraints on star formation theories.
Alternatively, BDs and isolated planetary mass objects
may be stellar ``embryos'' ejected from multiple
forming systems before reaching a stellar mass \citep{RC01}, or they may 
form like planets by coagulation of dust particles and
subsequent gas accretion within circumstellar protoplanetary disks
\citep{Lea98}. Young (proto)-planets could then be extracted or ejected by
dynamical interactions from the forming planetary systems. 
In
this scenario, isolated BDs and planetary mass objects
are intrinsically different from stars; their study sheds
no light on {\it star} formation theories, but provides instead a chance of
studying the early evolution of giant planets where they can be observed in
isolation, rather than very close to a much brighter star.

One way to shed light on the origin of very low mass objects is to
ascertain their association
with circumstellar disks, which are characteritics of stellar formation from the
contraction of a molecular core. Deep images in the L band \citet{Mea01} and in
the mid-infrared \citep{Pea00,Bea01} have shown that many  very low luminosity
objects have excess emission at these wavelengths, and detailed studies of some
of them have proven that the central objects are bona-fide BDs.
Their infrared excess is consistent with the 
presence of a surrounding disk similar to those found around more
massive pre-main sequence stars \citep{Cea98,Cea00,NT01}.
These initial findings, albeit limited, seem to suggest that indeed BDs 
form like ordinary stars. 

We report here the first results of a project
aimed on one hand
to improve our understanding of disks around BD, following the approach of
in \citet{NT01}, and on the other hand to find evidence of a
circumstellar disk around {\it bona fide} isolated planetary mass
objects. 
We have used the Near-Infrared Camera
and Spectrograph (NICS) on the 3.56m Telescopio Nazionale Galileo to acquire
low-resolution (NIR) spectra of a sample of mid-infrared sources
located within the $\rho$-Ophiuchi cloud \citet{Bea01}.  Our target list
included objects detected at 6.7 and 14.3~$\mu$m with the ISOCAM camera on
board of ISO \citep{Kea96,Cea96}, having low effective temperature
(T$_{eff}$), luminosity (L$_\ast$) and extinction (A$_{\rm V}$) based on
photometric or limited 2.2~$\mu$m spectroscopic estimates.
The goal of our new observations was to obtain improved
determinations of these parameters and to derive accurate values of masses and
ages of the targets by comparison with theoretical evolutionary tracks. 
The results for the complete sample will be discussed elsewhere 
\citep{Nea02}. In this Letter, we report on one of the sources,
number 33 of the ISOCAM list of \citet{Bea01}, which is associated with the 
NIR source GY11 \citep{GY92}. The substellar nature of this object
was already proposed by \citet{RR90} and confirmed by
\citet{Cea98,WGM99}. We derive a new, accurate spectral classification of GY11,
and of its effective temperature and luminosity. Our data suggest
that GY11 is a planetary mass object, with an
infrared excess that can be roughly modeled as due to a circumstellar disk
similar to those associated to T Tauri stars. These observations provide the
first evidence that objects of such small mass
actually form in a star-like manner, and thus that they are
genetically different from ``planets''.

\section{Observations}

A near-infrared low-resolution spectrum of GY11 was obtained with the
Telescopio Nazionale Galileo on La Palma
on July~9, 2001, using a 0\farcs5 slit and the
high-throughput low-resolution prism-based disperser unique to NICS
\citep{Baffa01},
the Amici device
\citep{O00}; this setup offers a complete NIR spectrum, 0.85 to 2.35~$\mu$m,
at R$\sim$100 accross the entire range, and it allows 
an accurate classification of faint and cool dwarfs \citep{Tea01}.
Instrumental and telluric correction was ensured by observations 
of A0 stars. The shape of the final spectrum was checked using 
near infrared photometric measurements from the 2MASS second incremental
data release; synthetic magnitudes were computed using the appropriate 
transmission curves and compared with the source photometry, colors were
found to be consistent with those of 2MASS to
within 10\%, as expected from typical uncertainties.
To better constrain the values of the
extinction, we also obtained optical i-band (0.77~$\mu$m) photometry at the
ESO-La Silla 1.54m Danish Telescope using the DFOSC instrument.
Photometric calibration was ensured
by observations of a set of \citet{L92} standard stars,
converted into the 
Gunn system using the transformations given by \citet{Fea96}.

\section{Central source parameters}


\citet{Bea01} associated the ISOCAM source~33 with a Class~II object member of
the $\rho$-Ophiuchi young stellar cluster known as GY11 \citep{GY92}. 
The brown
dwarf nature of GY11 has been suspected for some time \citep{RR90}. However,
there is a large uncertainty in the literature as to the exact value
of its photospheric parameters and mass. \citet{Bea01} 
estimate  bolometric luminosity and
extinction to be L$_\ast\sim 0.001$~L$_\odot$, A$_{\rm
V}$=2.7 mag from NIR photometry.  
Based on multiband infrared photometry and a 2.2$\mu$m low resolution
spectrum,  \citet{WGM99}
estimate a spectral type M6.5, A$_{\rm
V}\sim$5~mag, L$_\ast\sim$0.002~L$_\odot$, and T$_{eff}\sim$2650~K. These
authors noted that due to veiling caused by dust emission, the spectral type
could easily be  some 2--3
subclasses later, 
and the extinction significantly underestimated.  
In fact, \citet{Cea98} derive a higher value of A$_{\rm V}$=10~mag from
broadband photometric measurements that include optical and infrared bands.

Our complete near infrared spectrum offers the 
possibility of a better estimate of the source parameters, as it allow
us to use the global spectrum shape below 2~\um, a region which is less
affected by the continuum veiling due to the dust emission.
Given the expectation that the surface gravity of very young BDs shouls be 
similar to sub-giants, we derive the photospheric parameters by comparison
with field dwarfs spectra and model atmospheres with appropriate surface
gravity, as suggested by the evolutionary models \citep{Cea00}.
We first 
derive extinction and spectral type by matching the observed GY11 spectrum 
with that of field dwarfs in the solar neighborhood
\citep{Tea02}, obtained with the same
instrumental set-up and reddened using the 
\citet{Cea89} extinction law most appropriate
for Ophiucus (R$_V$=4.2; Fig.~1a). We try to provide the best fit to the 
 global shape of the spectrum, 
with  particular attention to
the shape of the H-band, the J-band features and 
the drop due to water vapor absorption at the red edge of the J-band. 
Overall, the best spectral match
is found with the field dwarf with spectral type M8.5
and extinction A$_{\rm V}\sim$7.0~mag.
Lower values of A$_{\rm V}$ (by $\sim 1$~mag)
offer a better match of the spectrum with later
dwarf spectra (M9--L0), but are not consistent with the broad band 
optical measurements (see inset in Fig.~4). 
A higher value of the extinction causes a too steep rise 
of the spectrum below 1.2~\um.
Field dwarfs with spectral types earlier than M7.5 show large deviations
from the observed shape of the H-band and
the drop at 1.3~$\mu$m.
Given the uncertainties of a classification based on objects 
with very different surface gravity, we expect our classification
to be accurate within one spectral class and the visual extinction
estimate within one magnitude.


As a second step, in order to derive an estimate of the photospheric
effective temperature (Fig.~1a), we compare
the observed GY11 spectrum,
with reddened, appropriate surface gravity,
$\log_{10} (g)$=3.5, model atmospheres 
\citep{Aea00,Chea00}.
The best estimate of the effective
temperature  is  $\sim$2400$\pm$100~K.
Higher temperature models offer a better match of the 
H-band shape, but underestimate the drop near 1.3~\um\ and the global
shape of the spectrum at J-band. The derivation of the effective
temperature of young dwarfs based on theoretical synthesis
of the near infrared spectrum is very uncertain
\citep{Lea02}; however,
our value of T$_{eff}$ for an object of spectral type M8.5
is consistent with the spectral type vs. effective temperature scale
discussed by \citet{WGM99}, and  only marginally
higher than the latest effective temperature scales derived for field dwarfs
\citep{Lea01}.

To estimate the luminosity (L$_\ast$) of the object, we used the 2MASS
J-band magnitude, dereddened  by  A$_{\rm V}\sim~ 7.0$ mag,
and the bolometric correction derived from the best fitting (2400~K)
atmospheric model.
The value of this ``theoretical'' bolometric correction is nearly identical
to the empirical value adopted by \citet{WGM99}. We derive a value of
L$_\ast\sim$0.008~L$_\odot\pm$30\%. 


Using the L$_\ast$ and T$_{eff}$ values derived above, we can compare 
the position of GY11 in the Hertzprung-Russell diagram with the predictions
of theoretical pre-main sequence evolutionary models. In Figure~2 we show 
this comparison for the latest release of  evolutionary tracks from  three
leading groups in the theory of pre-main sequence evolution of substellar 
objects. 
In spite of the relatively large uncertainties on
L$_\ast$ and T$_{eff}$, and on the limited accuracy of pre-main sequence 
evolutionary tracks at these ages and masses, we confirm that GY11 is a
young ($\tau<1$~Myr), very low mass object, probably below or very close to the
deuterium burning limit, with a best mass estimate in the range 7 to 12~M$_{Jupiter}$. 

\section{Infrared Excess and Disk Models}
GY11  is the lowest mass object with a clearly detected infrared excess.
It is  detected by ISOCAM
in the two broad-band filters LW2 and LW3 ($\lambda_{eff}$ 6.7 and
14.3 \um, respectively) used by \citep{Bea01} in
their $\rho$ Oph survey, as well as in the three intermediate-band filters,
SW1, LW1, LW4 ($\lambda_{eff}$=3.6, 4.5 and 6.0 \um, respectively),
used by \citet{Cea98} in their pointed observations of optically
identified candidate brown dwarfs. 
The ISOCAM measurements in the broad- and narrow-bands
are in good agreement, within the flux calibration
uncertainties, which we assume to be $\sim 20$\%. 

As usually with ISOCAM, the $\sim$6$^{\prime\prime}$ beam
includes multiple sources
seen in higher resolution
near-infrared images, which may contribute to the observed fluxes. 
Figure~3 compares
a K$_s$ image of the region around GY11
extracted from the VLT/ESO archive (originally observed as
part of ESO proposal 63.I-0691) and the ISOCAM LW1 ($\lambda_{eff}$=4.5 \um)
image from the ISO archive \citep{Cea98}. The mid-infrared emission peaks very
close to the position of GY11. Although  
a small contamination from the NIR source
$\sim$8$^{\prime\prime}$ to the east is possible,
we think that most of the mid-infrared observed flux comes  from GY11; a
similar conclusion was also reached by \citet{Cea98}.

%

In Figure~4,
we show the spectral energy distribution (SED) of GY~11 at all wavelengths
and compare it to that of an irradiated,
flared disk  similar to those that  reproduce the 
observed characteristics of TTauri systems \citep{CG97}.  The disk
has a dust mass of 1 Earth mass ($\sim$3\% of the mass of the central object,
for an assumed gas-to-dust mass ratio of 100) and
is heated by a
a central source with the GY11 temperature, luminosity and mass.
We show the predicted SED when the disk extends inward to the stellar surface
(solid line) and  when it has an inner hole of 3 stellar radii (dotted line).
More details on the disk models can be found in
\citet{NT01} and \citet{Nea02}.
The agreement between observed and predicted fluxes is rather good, 
especially for the disk with the large inner hole.
In particular, both models
predict total (star+disk) fluxes that
in the J, H, K bands are smaller than the calibrated TNG fluxes by
15\% at most.

As an independent check,
we computed  optical and NIR broad-band
magnitudes from the model-predicted SED.
They are compared in the inset of Figure~4 with the
observed dereddened magnitudes
in i (this paper), R,I,L$^\prime$ \citep{Cea98}, J,H,K (2MASS).
The agreement is again quite good.

The ISOCAM measurements, especially that at 14.3 \um, have large
error bars, and one should not overinterpret them. However,
is of some interest to point out that, if indeed the mid-infrared excess
is due to disk emission,
the disk must be flared. Therefore, 
dust and gas must be well mixed, as in the majority of pre-main--sequence
stars, and  no major settling of the dust onto the disk midplane has
occurred during the lifetime of GY11. The disk must be optically thick
to mid-infrared radiation; this however sets only a lower limit to the
disk mass of roughly 10$^{-5}$-10$^{-6}$ M$_\odot$ depending on the
exact value of the
dust mid-infrared opacity and surface density profile.  Note that the
disk mass has to be very small; if we assume the ratio of the disk
mass to the mass of the central object typical of TTS (~0.03), then the
disk mass is  about $3\times 10^{-4}$~M$_\odot$, and the disk contains only 1 Earth mass
of dust. As a consequence, the accretion rate (if any) is also likely
to be low, with  an average value over the lifetime of the object that
cannot exceed 3$\times$10$^{-10}$~M$_\odot$\,yr$^{-1}$
(for an age of 1~Myr). The accretion
luminosity is also small, about 40 times lower than the luminosity of the
photosphere. A direct determination of the disk mass can be derived from
(sub)millimeter wavelength observations. We predict  for GY11 a 1.3mm
flux of about 0.6 mJy, which is well below the upper limit set by
the survey of \citet{MAN98}, but within the expected capabilities
of the ALMA millimeter array.

\section{Conclusions}


The results presented in this letter show evidence that a  young isolated
planetary mass objects in
$\rho$-Oph, with mass of about 10 M$_{\rm J}$, is surrounded by warm dust,
possibly distributed on a disk similar in properties to those around  young
brown dwarfs and T Tauri stars.  The implications of this finding, that should
be  confirmed by higher spatial resolution mid-infrared observations, and
should be extended to a large sample of similar objects, are profound, since it
gives a clear indication that isolated BDs and even planetary mass objects
form like stars and are not produced in a
planet-like fashion within protoplanetary disks around more massive objects,
and later ejected by dynamical interactions. Isolated BDs and
planetary mass objects are thus an extension of the
stellar and substellar sequence to very low masses and have different origins
from ``planets''.

\acknowledgments
This work is partly based on observations collected at the Italian Telescopio 
Nazionale Galileo, Canary Islands, Spain, at the European Southern Observatory
telescopes on La Silla and Paranal observatories, Chile, and on data obtained
by the European Space Agency Infrared Space Observatory.
This publication makes use of data products from the Two Micron All Sky Survey,
which is a joint project of the University of Massachusetts and the Infrared
Processing and Analysis Center/California Institute of Technology, funded by
the National Aeronautics and Space Administration and the National Science
Foundation. This work was partly supported by  ASI grant  ARS 1/R/27/00 to the
Osservatorio di Arcetri.

%
%

\clearpage
\begin{figure}
\vspace{9.5cm}
\includegraphics{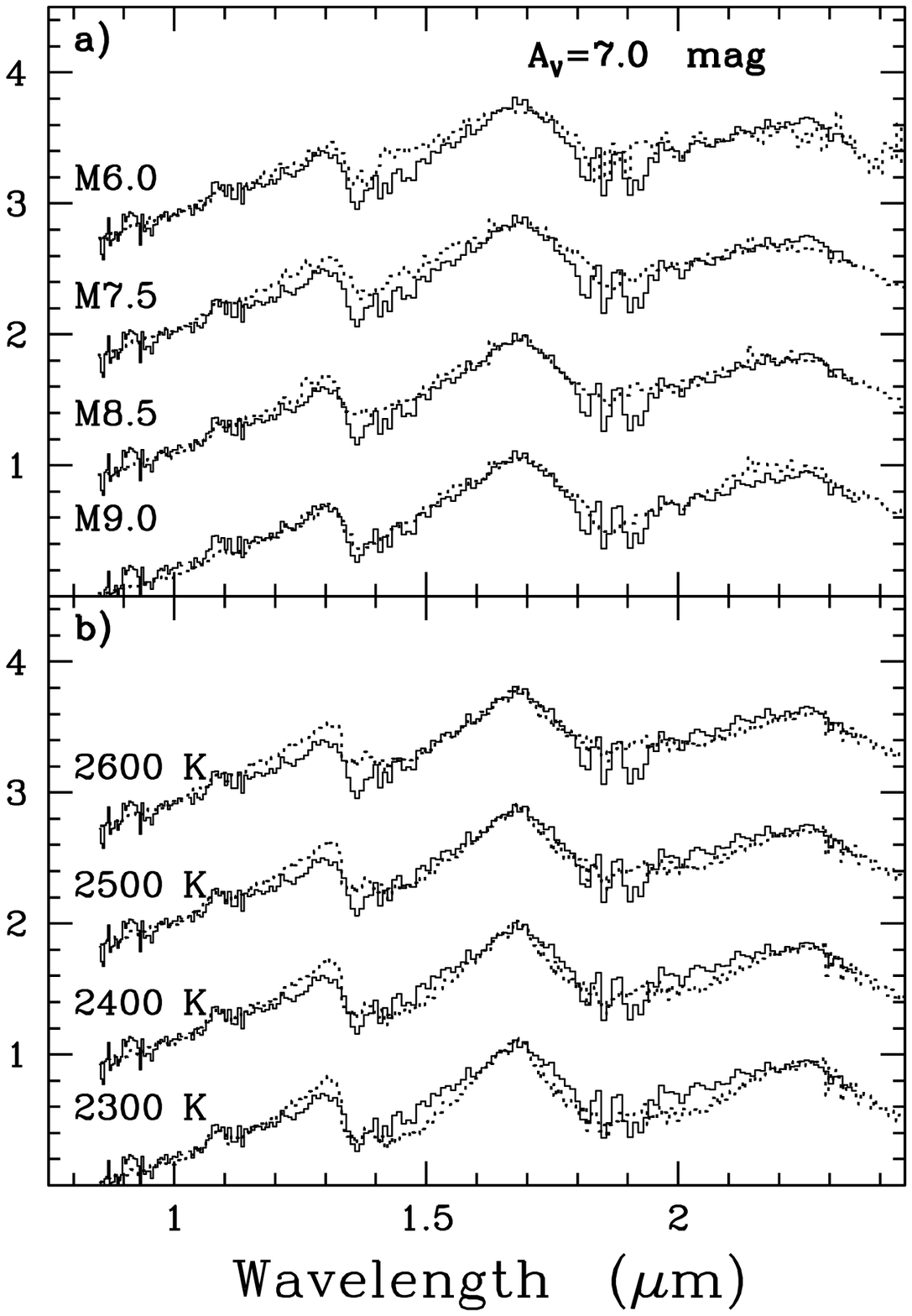}
\caption{The spectrum of GY11 (thick solid line)
is compared with reddened spectra of field M-dwarfs (dotted lines) as labelled
(from Testi et al.~2002), all spectra 
are normalized at the mean flux in the 1.6-1.7~$\mu$m range and shifted with
constant offsets for clarity. The spectrum of GY11 is reproduced at every offset
to ease the comparison. b) similar to a), but the dotted spectra are reddened
theoretical atmospheric models \citep{Aea00}, T$_{eff}$ as labelled.
In both panels, the observed spectra have a lower signal to noise in the region 
corresponding to the strong telluric absroption (1.35--1.45\,$\mu$m and
1.82--1.95\,$\mu$m). 
}
\end{figure}

\clearpage 
\begin{figure}
\vspace{7.5cm}
\includegraphics{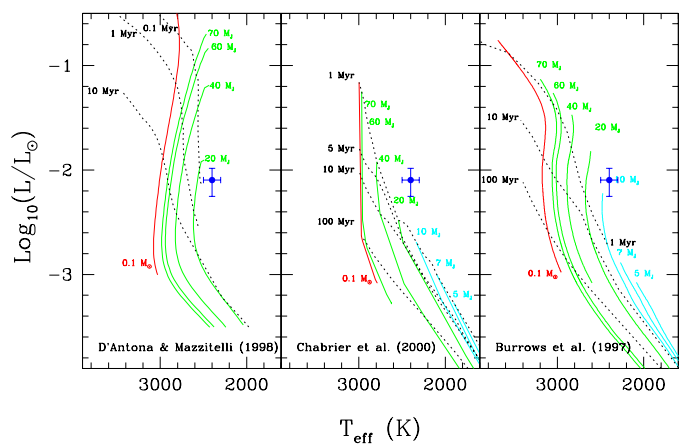}
\caption{Hertzprung-Russell diagram for three sets of
evolutionary tracks \citep{DM97,Chea00,Bea97}. The position of GY11 is shown
as a blue circle. The tracks are labelled with the appropriate mass,
hydrogen burning stars are shown in red,
deuterium burning brown dwarfs in green, and objects below the 
deuterium burning limit in cyan. Isochrones are shown as dotted
lines and are labelled with the appropriate ages.}
\end{figure}

%

\clearpage
\begin{figure}
\vspace{7.5cm}
\includegraphics{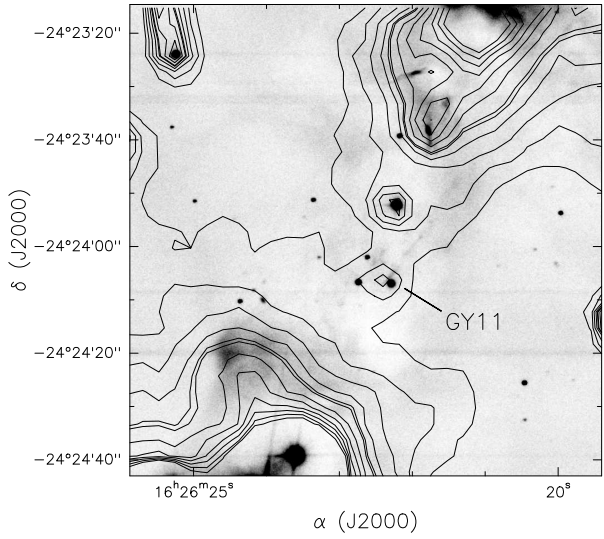}
\caption{The ISOCAM-LW1 (contours, 4.5$\mu$m) image of the
region surrounding GY11 is overlaid on the VLT K$_s$ (greyscale, 2.2$\mu$m)
image. The ISOCAM image has been aligned with the VLT image
by matching the position of GY10 (2320.8-1708, \cite{Cea98})
with the associated mid-infrared source.
}
\end{figure}

\clearpage
\begin{figure}
\vspace{7.5cm}
\includegraphics{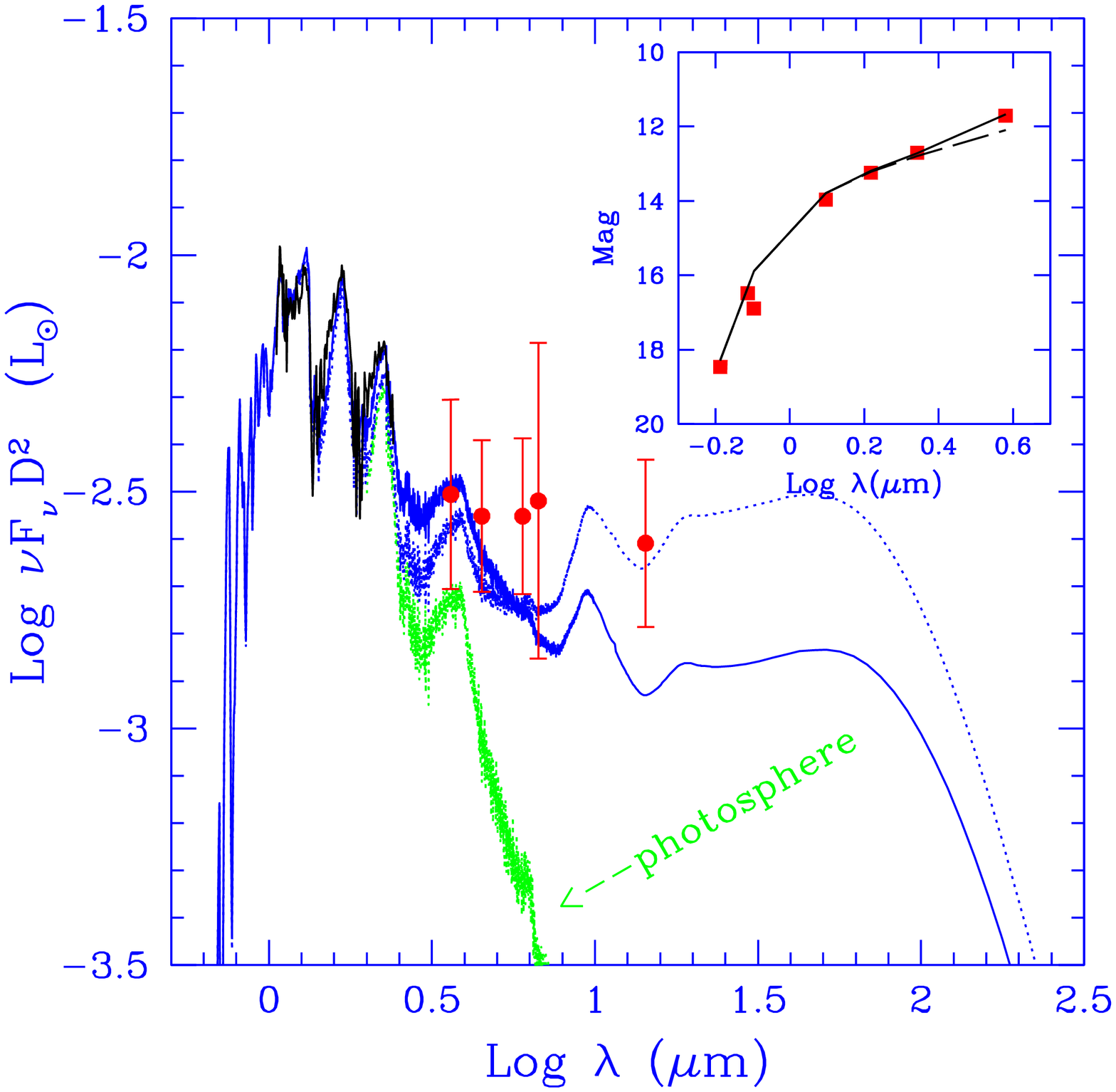}
\caption{Disk and photosphere model  of the GY11 SED.
The red circles with errorbars show the mid-infrared fluxes from ISOCAM
\citep{Cea98,Bea01}, the black line is our NIR dereddened Amici
spectrum. The green jagged line is the photospheric model for T$_{eff}$=2400~K
and Log(g)=3.5 \citep{Aea00}. The blue lines show the combined photosphere plus
disk emission computed as in the text. In one case (solid line) the disk inner radius is equal to R$_\star$, in the other (dotted line) to 3R$_\star$.
In the inset, we show the comparison 
between the dereddened broad band photometry (from R to L$^\prime$; red squares)
and the models (practically coincident), photosphere as a dashed line, photosphere plus disk as solid
line.
}
\end{figure}

\end{document}